\documentclass[iop]{emulateapj}

\begin{document}

\author{Juncheng Chen\altaffilmark{1}, Xiaofeng Wang\altaffilmark{1}, Mohan Ganeshalingam\altaffilmark{2,3},
Jeffrey M. Silverman\altaffilmark{2,4,5}, Alexei V. Filippenko\altaffilmark{2}, Weidong Li\altaffilmark{2,6},
Ryan Chornock\altaffilmark{2,7}, Junzheng Li\altaffilmark{1}, and Thea Steele\altaffilmark{2} }

\altaffiltext{1}{Physics Department and Tsinghua Center for
Astrophysics (THCA), Tsinghua University, Beijing, 100084, China;
cjc09@mails.tsinghua.edu.cn; wang\_xf@mail.tsinghua.edu.cn.}

\altaffiltext{2}{Department of Astronomy, University of California,
Berkeley, CA 94720-3411, USA.}

\altaffiltext{3}{Environmental Energy Technologies Division, Lawrence Berkeley National
Laboratory, 1 Cyclotron Road, Berkeley, CA 94720, USA.}

\altaffiltext{4}{Department of Astronomy, University of Texas, Austin, TX 78712, USA.}

\altaffiltext{5}{NSF Astronomy and Astrophysics Postdoctoral Fellow.}

\altaffiltext{6}{Deceased 12 December 2011.}

\altaffiltext{7}{Harvard-Smithsonian Center for Astrophysics, 60 Garden
Street, Cambridge, MA 02138, USA.}

\title{Optical Observations of the Type Ic Supernova 2007gr in NGC 1058 \\ and Implications for the Properties of its Progenitor}

\begin{abstract}
We present extensive optical observations of the normal Type Ic supernova (SN) 2007gr, spanning from about one week before maximum light to more than one year thereafter. The optical light and color curves of SN 2007gr are very similar to those of the broad-lined Type Ic SN 2002ap, but the spectra show remarkable differences. The optical spectra of SN 2007gr are characterized by unusually narrow lines, prominent carbon lines, and slow evolution of the line velocity after maximum light. The earliest spectrum (taken at $t=-8$ days) shows a possible signature of helium (He~I $\lambda$5876 at a velocity of $\sim$\,19,000\,km\,s$^{-1}$). Moreover, the larger intensity ratio of the [O~I] $\lambda$6300 and $\lambda$6364 lines inferred from the early nebular spectra implies a lower opacity of the ejecta shortly after the explosion. These results indicate that SN 2007gr perhaps underwent a less energetic explosion of a smaller-mass Wolf-Rayet star ($\sim 8$--9\,M$_{\odot}$) in a binary system, as favored by an analysis of the progenitor environment through pre-explosion and post-explosion {\it Hubble Space Telescope} images. In the nebular spectra, asymmetric double-peaked profiles can be seen in the [O~I] $\lambda$6300 and Mg~I] $\lambda$4571 lines. We suggest that the two peaks are contributed by the blueshifted and rest-frame components. The similarity in velocity structure and the different evolution of the strength of the two components favor an aspherical explosion with the ejecta distributed in a torus or disk-like geometry, but inside the ejecta the O and Mg have different distributions.

\end{abstract}


\section{Introduction}
Supernovae (SNe) are classified into two main categories based on their explosion mechanisms: thermonuclear and core collapse (CC). Stellar explosions of the former kind are Type Ia supernovae (SNe~Ia), characterized by a deep blueshifted Si~II $\lambda$6355 absorption trough in their optical spectra, and are generally believed to result from the thermonuclear runaway of an accreting carbon-oxygen white dwarf (WD) in a binary system (Wang et al. 2013, and references therein). The latter variety, core-collapse explosions of massive stars, can be further subdivided as Type II, Ib, and Ic SNe in terms of diverse appearances of the hydrogen and/or helium features in their optical spectra (Filippenko 1997). SNe~II are defined by the presence of prominent hydrogen lines in the spectra, while SNe~Ib and SNe~Ic do not exhibit these lines. The presence of noticeable optical He~I lines, especially He~I $\lambda$5876, distinguishes SNe~Ib (He-rich) from SNe~Ic (He-poor); see Wheeler \& Harkness (1986), Harkness \& Wheeler (1990), and Filippenko, Porter, \& Sargent (1990). Such a wide variety of spectroscopic properties is related to the stripping of the outer layers of their progenitor stars, which depends on their initial mass and environments (single or binary, separation, etc.; e.g., Pols \& Nomoto 1997).

The absence of H and He lines in SN~Ic spectra indicates that their progenitors have lost the outer envelope of hydrogen and most of the helium before explosion. Outer layers of stars can be shed through strong radiatively driven stellar winds from the progenitor star itself, or are stripped off by a companion star (Podsiadlowski et al. 1992). The progenitors can thus be restricted to either relatively low-mass stars in a binary system or single massive stars with main-sequence masses of 30--40\,M$_{\odot}$ such as Wolf-Rayet stars (Wheeler \& Levreault 1985). However, direct detections of SN~Ic progenitors still proves elusive (Smartt et al. 2009; Eldridge et al. 2013). The presence of moderately strong Si~II lines in their spectra make SNe~Ic and SNe~Ia superficially similar at early times, possibly contaminating the SN~Ia sample at higher redshifts when using low-quality spectra.

Moreover, SNe~Ic have gained special attention in the past few decades because of their relation to gamma-ray bursts (GRBs). Some energetic, broad-lined (BL) SNe~Ic have been associated with long-duration GRBs at relatively low redshifts. Well-studied cases include (among others) SN 1998bw--GRB 980425 (Iwamoto et al. 1998), SN 2003dh--GRB 030329 (Matheson et al. 2003), and SN 2010bh--GRB 100316D (Bufano et al. 2012). Besides these BL~SNe~Ic, very few observations and investigations have been carried out for normal or narrow-lined (NL) SNe~Ic. Better studies of these objects in different subclasses will help reveal and explain the observed diversity of stripped-envelope SNe.

SN 2007gr was discovered at a relatively young phase in the nearby galaxy NGC 1058 (Madison \& Li 2007). Based on a spectrum obtained the following night, SN 2007gr was initially classified (Chornock et al. 2007) as a generic SN~Ib/c since it was not clear whether the line near 6350\,\AA\ was really He~I (as for SNe~Ib). Later spectra did not reveal significant He, and the object was reclassified as a SN~Ic (Valenti et al. 2008b, hereafter V08).

Owing to the brightness of SN 2007gr, extensive multi-band observations of it were conducted immediately after the discovery. The earlier optical spectra show that it has unusually narrow lines and prominent carbon features, though the peak luminosity is normal (V08). The narrow spectral lines indicate a lower kinetic energy per unit mass of the ejecta. By modeling the nebular spectra of SN 2007gr, Mazzali et al. (2010) favored a low-mass progenitor star and an explosion with low kinetic energy, consistent with the results from the radio and X-ray observations (Soderberg et al. 2010; but see Paragi et al. 2010). Pre-explosion {\it Hubble Space Telescope (HST)} images of SN 2007gr reveal that it is possibly located within a compact cluster having an age of $\sim 7$\,Myr or 20--30\,Myr, suggesting an initial mass of $28\pm4$\,M$_{\odot}$ or 9--12\,M$_{\odot}$, respectively, for the progenitor star (Crockett et al. 2008).

In this paper, we present extensive observations of SN 2007gr in optical bands, spanning from about 8 days before maximum light to $\sim 450$ days thereafter, providing an excellent sample for comparison with other SNe~Ib/c. Hunter et al. (2009, hereafter H09) and V08 have previously studied the optical properties of SN 2007gr, but our dataset provides an independent sample having slightly longer temporal coverage, at early and late phases, and therefore can help better constrain the object's properties. Moreover, we can put stronger constraints on the progenitor of SN 2007gr by taking advantage of pre- and post-explosion data obtained with {\it HST}. Our observations and data-reduction methods are described in \S 2. In \S 3 we present the optical light curves and color curves. The quasi-bolometric light curve constructed by integrating the optical fluxes is given in \S 4. In \S 5 we described the spectral evolution. Our discussion and conclusions are given in \S 6.

\section{Observations and Data Reduction}

SN 2007gr was discovered (Madison \& Li 2007) on 2007 August 15.51 (UT dates are used throughout this paper) by the Lick Observatory Supernova Search (LOSS) using the 0.76\,m Katzman Automatic Imaging Telescope (KAIT; Filippenko et al. 2001). Its J2000 coordinates are $\alpha = 02^h43^m30^s.39$, $\delta{} = +37^\circ{}20'27''.4$, and it exploded $24.8''$ west and $15.8''$ north of the nucleus of NGC 1058; see Figure 1. Its host is a nearby spiral galaxy that previously harbored SN 1969L and SN 1961V. Extensive follow-up observations of SN 2007gr were triggered immediately after the discovery.

\begin{figure}
\begin{center}
\includegraphics[angle=0,width=0.5\textwidth]{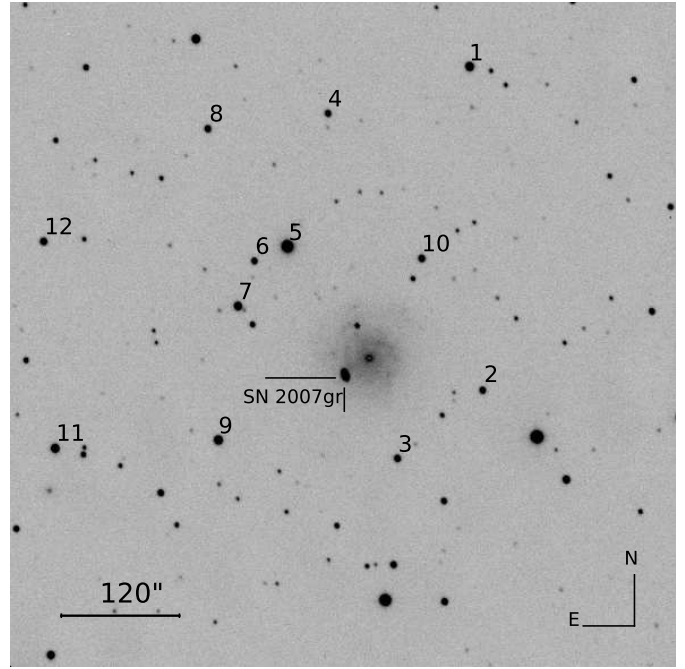}
\end{center}
\caption{\footnotesize
SN 2007gr in NGC 1058. This is a $V$-band image taken with the 0.8\,m TNT on 2007 September 9. The supernova (which is slightly northeast of a foreground star) and 12 local comparison stars are marked.}
\label{fig:field}
\end{figure}

\subsection{Ground-Based Observations}

Ground-based optical ($UBVRI$) photometry of SN 2007gr was obtained with the 0.76\,m KAIT, the 1.0\,m Nickel telescope at Lick Observatory, and the 0.8\,m THU-NAOC Telescope (TNT) at Beijing Xinglong Observatory (BAO) in China (Wang et al. 2008; Huang et al. 2012). The instrumental response functions of the former telescopes, obtained by multiplying the filter transmission functions by the quantum efficiency of the CCD detectors and the atmospheric transmission, can be seen in Figure 2 of Wang et al. (2009). The response curves of the TNT are not shown because the accurate transmission curves of the $UBVRI$ filters are not available.

All of the KAIT and Nickel images were reduced using a mostly automated pipeline (Ganeshalingam et al. 2010). The TNT images were reduced with the Image Reduction and Analysis Facility (IRAF\footnote[1]{IRAF, distributed by the National Optical Astronomy Observatory, which is operated by the Association of Universities for Research in Astronomy (AURA), Inc. under cooperative agreement with the National Science Foundation (NSF).}) standard procedures, including correction for bias, flat-field division, and removal of cosmic rays. We applied host-galaxy subtraction to minimize contamination from the galaxy background and close field stars. Template images of the host galaxy were obtained when SN 2007gr had faded away with negligible flux. The KAIT, Nickel, and TNT templates of the host galaxy were obtained on 2010 December 2, 2008 August 30, and 2010 November 29, respectively. The point-spread function (PSF) fitting method was applied to obtain the instrumental magnitudes of SN 2007gr and 12 local standard stars as shown in Figure 1.

Instrumental magnitudes were then converted into the standard Johnson $UBV$ (Johnson 1966) and Kron-Cousins $RI$ (Cousins 1981) systems, with the color terms and zeropoints determined by measuring a series of Landolt (1992) standards on the photometric nights. Table 1 lists the average color terms for the relevant telescopes and filters. The $UBVRI$ magnitudes and uncertainties of the local standard stars are listed in Table 2. To alleviate systematic discrepancies in the photometric systems on different telescopes, we further applied {\it S}-corrections (Stritzinger et al. 2002) to our photometry using our own spectra presented in this paper as well as those published by V08. After the {\it S}-corrections, the KAIT and Nickel photometry agree to within 0.05 mag. The {\it S}-corrections were not applied to the TNT data owing to the lack of accurate response curves of the corresponding filters. The final flux-calibrated and {\it S}-corrected $UBVRI$ magnitudes are presented in Table 3, and the corresponding light curves are shown in Figure 2.

\begin{figure}
\begin{center}
\includegraphics[angle=0,width=0.5\textwidth]{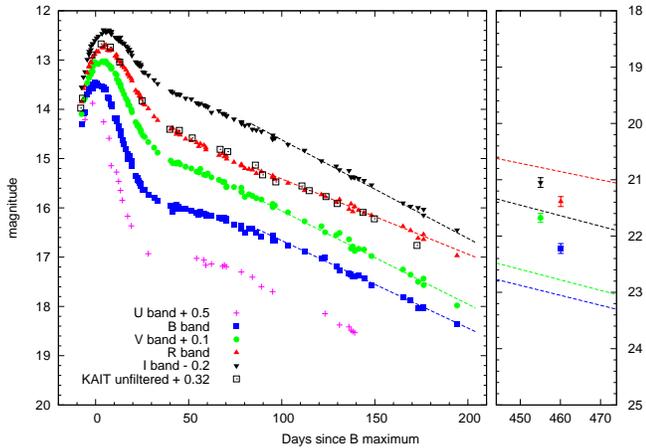}
\end{center}
\caption{\footnotesize
$UBVRI$ light curves of SN 2007gr, shifted for clarity. Very late-time {\it HST} photometry is plotted in the right panel. The dashed lines show linear fits to the late-phase ground-based photometry. }
\label{fig:otlc}
\end{figure}

\subsection{{\it HST} Late-Time Photometry}

Two late-time images of SN 2007gr were obtained with the {\it HST}/{\it Wide Field and Planetary Camera 2} (WFPC2) on 2008 Nov. 20 and 25 (GO-10877; PI W. Li), corresponding to $t \approx 455$ days after maximum light. The observations were made in the F450W, F555W, F675W, and F814W filters, for which the transmission curves are very similar to those of the Johnson-Cousin $BVRI$ filters, as illustrated by the small coefficients required for the transformations between these two systems (Sirianni et al. 2005). The total exposure times were 800, 460, 360, and 700\,s, respectively. Photometry was extracted using the DOLPHOT 2.0 WFPC2 module, which is an adaptation of HSTphot (Dolphin 2000). With the spectrum taken around 430 days after $B$-band maximum, we apply {\it S}-corrections to convert the magnitudes of the {\it HST} system into those of the Johnson-Cousin system. The final magnitudes are listed in Table 3.

\subsection{Optical Spectroscopy}

There are 13 low-resolution optical spectra of SN 2007gr obtained with the Kast double spectrograph (Miller \& Stone 1993) on the 3\,m Shane telescope at Lick Observatory and the Low Resolution Imaging Spectrometer (LRIS; Oke et al. 1995) on the 10\,m Keck-I telescope, spanning from $-7.8$ to $+430.2$ days since $B$-band maximum. A journal of spectroscopic observations is given in Table 4.

All data were reduced with standard IRAF routines and IDL scripts described by Silverman et al. (2012). The spectra were flux calibrated using standard stars observed during the same night at airmasses similar to those of the SN. The slit was aligned along the parallactic angle for each of our spectral observations to minimize the chromatic effects of atmospheric dispersion (Filippenko 1982), but in any case the Keck/LRIS spectra were taken with an atmospheric dispersion compensator. Absolute fluxes of the spectra were not checked against the photometry, but the relative spectrophotometry is accurate: the colors synthesized from spectra agreed with the observed ones to within 0.05 mag (see also the discussion by Silverman et al. 2012). The spectral evolution of SN 2007gr is displayed in Figure 3. All spectra were corrected for the host-galaxy redshift of $z=0.001728$ (Tifft \& Cocke 1988).

\begin{figure}
\begin{center}
\includegraphics[angle=0,width=0.5\textwidth]{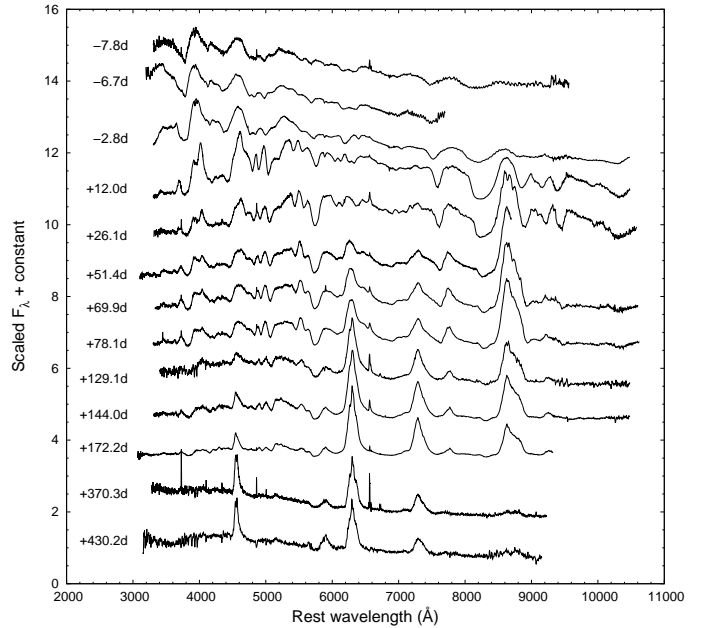}
\end{center}
\caption{\footnotesize
Optical spectral evolution of SN 2007gr. The spectra have been corrected for the host-galaxy redshift of 0.001728 (Tifft \& Cocke 1988),
as well as scaled and shifted vertically for clarity. The numbers on the left-hand side mark the epochs of the spectra in days since
$B$-band maximum. Note that the very narrow emission lines visible in some of the spectra are produced by H~II regions in the host galaxy.}
\label{fig:otspec}
\end{figure}

\section{Optical Light Curve and Color Evolution}

In Figure 2, we present the $UBVRI$ light curves of SN 2007gr. The very late-time photometry obtained with {\it HST} WFPC2 at $t \approx 455$ days is also shown in the right panel of the plot.

The peak magnitudes and corresponding dates in different bands were estimated by using a parabolic fit to the observed light curves around maximum brightness. The best-fit parameters are listed in Table 5. As for SNe~Ia, we estimated the initial magnitude decline rate in $B$ after maximum light (dubbed $\Delta m_{15}(B)$; Phillips 1993) for SN 2007gr: $1.31 \pm 0.04$ mag. The light curves generally exhibit a linear (in magnitudes) decline in the nebular phase, but with a break at $t \approx 80$ days; the $UBVI$ light curves decline faster thereafter. This transition can be attributed primarily to an increase in the escape rate of $\gamma$-ray photons. However, the $R$-band light curve seems to exhibit a slower decay at this time, likely owing to the contribution of [O~I], [O~II], and [Ca~II] emission in the nebular phase (see also Fig. 8). The very late-time {\it HST} photometry in $B$ and $V$ is obviously brighter than the linear fits to the late-time ground-based light curves (dashed lines in Fig. 2). This may be caused by the radioactive decay of long-lived isotopes, interaction with the circumstellar medium (CSM), light echoes, and delayed optical input through recombination. Determination of the exact physics for this lingering light is beyond the scope of this paper.

In Figure 4, we compare the $UBVRI$ light curves of SN 2007gr with those of well-observed SNe~Ib/c such as SN 1994I (Filippenko et al. 1995; Richmond et al. 1996), SN 2002ap (Foley et al. 2003), SN 2004aw (Taubenberger et al. 2006), SN 2008D (Modjaz et al. 2009; Tanaka et al. 2009),
and the normal SN~Ia 2005cf (Wang et al. 2009). The light curves of these SNe have been shifted in magnitude and phase to match the maximum brightness of SN 2007gr in different bands. The observations of SN 2007gr from H09 are also plotted. Our results are generally consistent with those of H09. The small deviations seen in the $U$ band at late times likely occur because the TNT data were not {\it S}-corrected owing to the lack of accurate response curves of the corresponding filters.

One can see from Figure 4 that the light curves of SNe~Ib/c exhibit much diversity. SN 2007gr has a post-maximum decline rate very similar to that of SN 2002ap. Their decline rates are slower than that of SN 1994I but faster than those of SN 2008D and SN 2004aw. We further notice that such differences are larger in $V$ and $R$ but smaller in $U$ and $B$. The origin of this wavelength dependence is probably related to the evolution of spectral features (see \S 5). Compared to the standard SN Ia 2005cf, the light curves of SNe~Ib/c do not show the secondary peak in the $R$ and $I$ bands. Moreover, the decline rate in the nebular phase for SNe~Ib/c seems to be smaller than that of a typical SN~Ia; e.g., 1.11 mag (100 day)$^{-1}$ for SN 2007gr and 1.62 mag (100 day)$^{-1}$ for SN 2005cf in the $B$ band. These differences are related to the mass of ejecta and iron-group elements produced in the explosion. Relative to SNe~Ia, the ejecta mass in SNe~Ib/c is usually larger and the synthesized nickel mass is usually smaller. Thus, SNe~Ib/c have a slower decay rate because the gamma-ray photons are largely trapped within the ejecta as a result of the larger opacity. On the other hand, the secondary maximum of the Type Ia SNe is proposed to be related to the ionization of iron-group elements in the ejecta, with a larger mass of $^{56}$Ni (and hence a brighter SN) producing a more prominent secondary peak (Kasen 2006). This may account for the absence of a secondary maximum in SNe~Ib/c because only a small amount of radioactive nickel is synthesized during the explosion.

\begin{figure}
\begin{center}
\includegraphics[angle=0,width=0.5\textwidth]{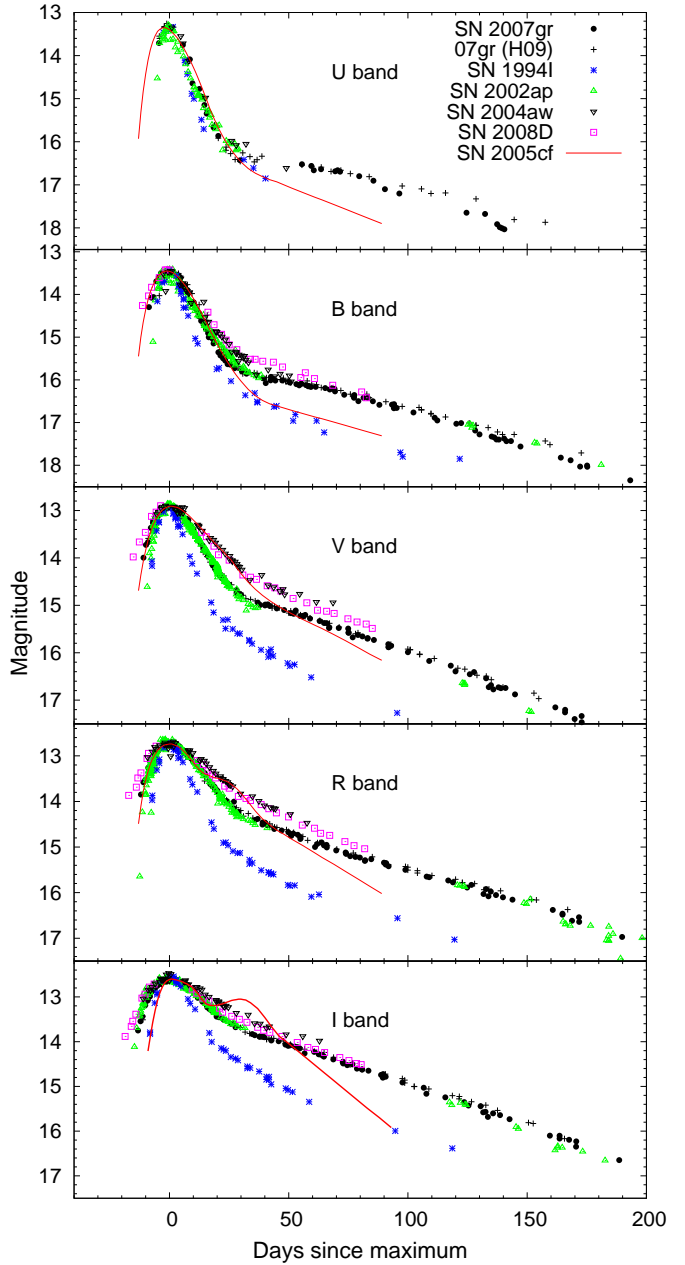}
\end{center}
\caption{\footnotesize
Comparison of the $UBVRI$ light curves of SN 2007gr with those of other SNe~Ib/c: SN 1994I (Filippenko et al. 1995), SN 2002ap (Foley et al. 2003), SN 2008D (Modjaz et al. 2009; Tanaka et al. 2009), and SN 2004aw (Taubenberger et al. 2006). The plus symbols represent data from H09. Light curves of SN~Ia 2005cf (Wang et al. 2009) are also presented for comparison. The light curves have been normalized as described in the text.}
\label{fig:com_lc}
\end{figure}

The color curves of SN 2007gr and the comparison sample are shown in Figure 5, all corrected for reddening in both the Milky Way and host galaxies.
The extinction for SN 2007gr is discussed in \S 4, and the values for the comparison SNe are obtained from the literature.
It is clear that different SNe~Ib/c exhibit similar color evolution at early phases, becoming progressively redder around maximum light and peaking at $t \approx 20$ days. However, significant differences emerge at $t>20$ days except in $V-I$, where a plateau is maintained until $t \approx 150$ days after maximum light. The large scatter at this phase is caused primarily by the abnormal behavior of SN 1994I, which exhibits the fastest post-maximum decline (especially at longer wavelengths). After the peak, the $B-V$ color becomes progressively bluer during later phases. The $V-R$ color has a more complicated evolution, with an initial decline toward the blue and a later reverse toward red values, perhaps because [O~I] emission gains strength during this phase. Similar evolution may exist for the other comparison SNe. Relative to the $B-V$ and $V-R$ colors, $V-I$ exhibits more uniform evolution among our sample; it maintains a constant value of $0.9 \pm 0.1$ mag from $t \approx 20$ days to $t \approx 150$ days after maximum light. Such a feature may be potentially used to estimate the reddening of SNe~Ib/c (see also the use of $V-R$ color as an extinction indicator, proposed by Drout et al. 2011). In comparison with normal SNe~Ia such as SN 2005cf (Wang et al. 2009), SN 2007gr and other SNe~Ib/c have much redder colors, and they also show different color evolution at late phases (especially in $V-R$ and $V-I$). For example, the $V-R$ color of the SN Ia 2005cf does not evolve toward the red after $t=80$ days past maximum light, and the $V-I$ color of SN 2005cf does not show the plateau feature seen in SNe~Ib/c.

\begin{figure}
\begin{center}
\includegraphics[angle=0,width=0.5\textwidth]{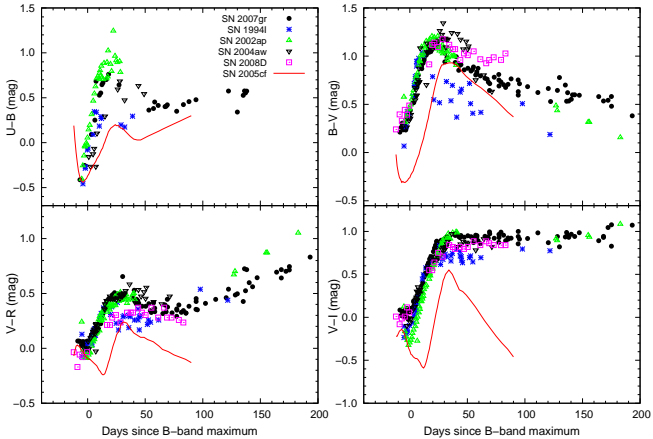}
\end{center}
\caption{\footnotesize
$U-B$, $B-V$, $V-R$, and $V-I$ color curves of SN 2007gr, compared with those of other SNe~Ib/c: SN 1994I (Filippenko et al. 1995), SN 2002ap (Foley et al. 2003), SN 2008D (Modjaz et al. 2009; Tanaka et al. 2009), and SN 2004aw (Taubenberger et al. 2006). Color curves of SN~Ia 2005cf (Wang et al. 2009) are also presented for comparison. All of the color curves have been dereddened with the values quoted in the references.}
\label{fig:col}
\end{figure}

\section{Absolute Magnitude and Bolometric Light Curve}

In calculating the absolute magnitudes and bolometric light curve, we adopted a distance modulus of $30.13 \pm 0.35$ mag for NGC 1058 (Schmidt et al. 1994). The Galactic reddening toward SN 2007gr is $E_{B-V}=0.055$ mag (Schlafly \& Finkbeiner 2011). The host-galaxy reddening is estimated to be $E_{B-V} = 0.030 \pm 0.023$ mag by using the empirical relation between the equivalent width of Na~I $\lambda$5890 and host color excess (Poznanski, Prochaska, \& Bloom 2012). We can also estimate the host-galaxy reddening based on a photometric method proposed by Drout et al. (2011), who found that the $V - R$ color of extinction-corrected SNe~Ib/c is tightly clustered at $0.26 \pm 0.06$ mag at $t \approx 10$ days after $V$-band maximum and $0.29 \pm 0.08$ mag at $t \approx 10$ days after $R$-band maximum. Applying this empirical relation to the observed $V-R$ color of SN 2007gr (which is estimated to be $0.32 \pm 0.04$ and $0.40 \pm 0.04$ at the above times, respectively) yields a host-galaxy reddening of $E_{B-V} = 0.11 \pm 0.08$ mag for SN 2007gr. In this paper we adopt the result derived from the equivalent width of the Na~I line, since the photometric method suffers a larger uncertainty. The Galactic and host-galaxy extinctions were obtained assuming $R_{V} =3.1$ (e.g., Cardelli et al. 1989). The absolute peak magnitudes of SN 2007gr are presented in Table 5.

To better quantify the explosion energy of SN 2007gr, we constructed its quasi-bolometric light curve using the $UBVRI$ photometry presented in this paper. In the calculation, we adopted the normalized passband transmission curves given by Bessell (1990). The integrated flux in each filter was approximated by the mean flux multiplied by the effective width of the passband. The missing data at some epochs were interpolated by using the neighboring measurements. Figure 6 shows the reddening-corrected, quasi-bolometric ($UBVRI$) light curves of SN 2007gr and several other SNe~Ib/c. The dashed line shows the decay rate from cobalt to iron.

Of the SN~Ib/c sample selected for comparison, SN 2004aw is the most luminous and exhibits the slowest post-maximum decline. SN 1994I has the narrowest light curve, but it has the second-highest luminosity among our sample. The luminosity of SN 2007gr is very close to those of SN 2002ap and SN 2008D, while its rise time is longer than that of SN 2002ap but shorter than that of SN 2008D. Despite the similarity in the light-curve shapes of SNe~Ic, their peak luminosities exhibit large scatter, and no obvious relation between the light-curve shape and luminosity at maximum light can be found for our sample. This confirms that SNe~Ib/c do not bear a width-luminosity relation as seen in SNe~Ia. However, the uncertainties in luminosity produced by uncertainties in distance modulus and reddening are $\sim 20\%$ of the peak values, comparable to the luminosity dispersion. Note that Drout et al. (2011) also did not find a width-luminosity relation in a sample of 25 SNe~Ib/c.

\begin{figure}
\begin{center}
\includegraphics[angle=0,width=0.5\textwidth]{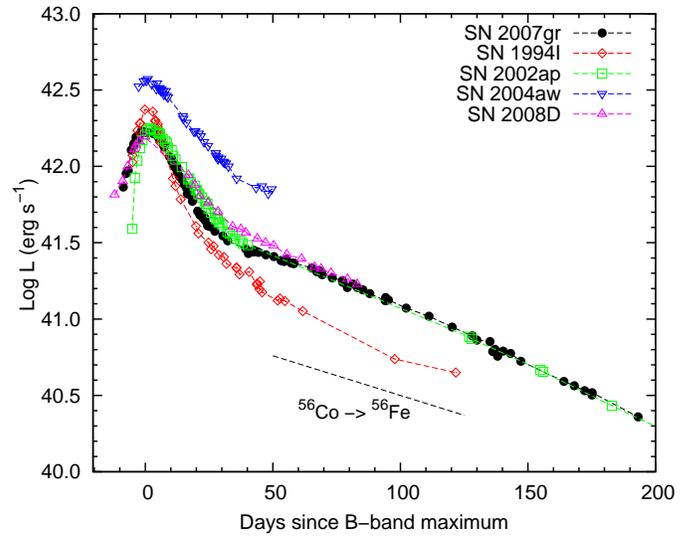}
\end{center}
\caption{\footnotesize
Quasi-bolometric ($UBVRI$) light curve of SN 2007gr, compared with those of SNe 1994I (Filippenko et al. 1995), SN 2002ap (Foley et al. 2003), SN 2008D (Modjaz et al. 2009; Tanaka et al. 2009), and SN 2004aw (Taubenberger et al. 2006).
}
\label{fig:BLC}
\end{figure}

We further estimated the mass of $^{56}$Ni synthesized during the explosion. Following the law of Arnett (1982), the maximum luminosity produced by the radioactive decay of $^{56}$Ni can be written as (Stritzinger \& Leibundgut 2005)
\begin{eqnarray}
M_{\rm Ni} = L_{\rm max} / ( 1.45 e^{-t_r/(111.3\,{\rm d})} + 6.45e^{-t_r/(8.8\,{\rm d})}),
\end{eqnarray}
where $t_r$ is the rise time of the bolometric light curve and $M_{\rm Ni}$ is the $^{56}$Ni mass (in units of solar mass, M$_{\odot}$). SN 2007gr was not detected in a KAIT image taken on 2007 Aug 10.44 ($\sim 5$ days before the discovery date) with a limit of $\sim 19.0$ mag. Assuming an explosion date in the middle of this interval, we find a rise time to maximum of $12.3 \pm 2.5$ days. Inserting this value and the maximum bolometric luminosity into the above equation, a $^{56}$Ni mass of $0.061 \pm 0.014$\,M$_{\odot}$ was derived for SN 2007gr. This value is slightly smaller than that obtained by H09 ($0.076 \pm 0.020$\,M$_{\odot}$), who used both optical and near-infrared (NIR) data in their calculations. For SNe~Ib/c, the contribution of NIR and ultraviolet emission to the total bolometric luminosity near maximum light can be up to $30\%$ (Stritzinger et al. 2009; Modjaz et al. 2009). Considering this difference, our estimate of the $^{56}$N mass is consistent with the results of H09.

\section{Spectroscopy}

A total of thirteen optical spectra of SN 2007gr were obtained with the Lick 3\,m Shane telescope and with the Keck-I telescope, spanning from $t = -7.8$ to $t = +430.2$~d with respect to the $B$ maximum. The complete spectral evolution is displayed in Figure 3. The main feature of the spectral evolution is that emission lines of [O~I] $\lambda$6300 and [O~II]+[Ca~II] at 7300\,\AA\ develop when entering the nebular phase. The line identifications and spectral evolution are discussed in the following subsections.

\subsection{Line Identifications in Early-Time Spectra}

In an attempt to identify the ions responsible for the lines in the spectra of SN 2007gr and to constrain the corresponding parameters such as the photospheric velocity and temperature, we adopted the parameterized resonance scattering synthetic-spectrum code SYNOW (Fisher et al. 1999; Branch et al. 2005). Figure 7 shows our best fit to the two spectra taken at $t = -7.8$ days and $t = +12$ days.

\begin{figure}
\begin{center}
\includegraphics[angle=0,width=0.5\textwidth]{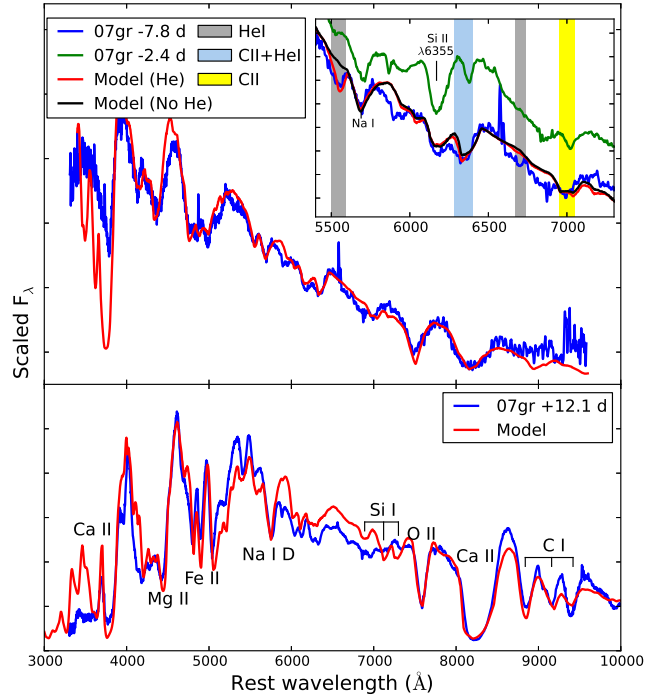}
\end{center}
\caption{\footnotesize
SYNOW fit to the spectra of SN 2007gr at $t=-7.8$ and 12.1 days relative to $B$-band maximum. The spectrum taken at $-2.8$ days is plotted in the inset of the upper panel for comparison. The best-fitting parameters are described in the text.}
\label{fig:Sfit}
\end{figure}

To fit the spectrum taken at $t=-7.8$ days (Fig. 7a), we assumed a photospheric velocity of $v_{\rm phot} =$ 10,000\,km\,s$ ^{-1}$ and a blackbody temperature of $T_{\rm bb} = 9000$\,K. One can see that the SYNOW fit matches well the observed spectrum, with the main absorption features at $\sim 4300$\,\AA\ (Mg~II $\lambda$4481), 4900\,\AA\ (Fe~II $\lambda\lambda$4924, 5018, 5169), 6200\,\AA\ (Si~II $\lambda$6355), 7600\,\AA\ (O~I $\lambda$7774), and 8200\,\AA\ (Ca~II $\lambda\lambda$8542, 8662, 8498) being well reproduced. The small notches at $\sim 6400$\,\AA\ and 7050\,\AA\ can be attributed to C~II $\lambda$6580 and C~II $\lambda$7235, respectively. Such identifications agree with those of V08 based on their spectrum. Note that the Fe~II $\lambda\lambda$4924, 5018, 5170 features are clearly separated, which is very uncommon among SNe~Ic.

There is a small notch at $\sim 5500$\,\AA\ in the spectrum at $t = -7.8$ days  that can be attributed to He~I $\lambda$5876 at a velocity of 17,000\,km\,s$^{-1}$. Another possible signature of helium is the weak absorption feature at $\sim 6700$\,\AA, which might be produced by He~I $\lambda$7065 at a velocity of 16,000\,km\,s$^{-1}$. Owing to blending with C~II $\lambda$6580, He~I $\lambda$6678 is hard to identify. The SYNOW fit indicates that the helium layer has a velocity of about 19,000\,km\,s$^{-1}$ and likely lies far above the photosphere. Possible helium features evolve rapidly and become nearly invisible in the spectrum at $t=-2.8$ days (see the inset in Fig. 7a). The detection of helium features in very early-time spectra of SN 2007gr suggests that the progenitor had a thin helium envelope immediately before the explosion.

Figure 7b shows the SYNOW fit to the $t=12$ day spectrum, with the adopted $v_{\rm phot}$ and $T_{\rm bb}$ being 7000\,km\,s$^{-1}$ and 7400\,K, respectively. It is apparent that the strengths of some lines of ionized species, such as Si~II and C~II, become weaker with decreasing photospheric temperature, while the corresponding neutral lines emerge at this phase. In particular, there are rather strong absorption features associated with C~I lines in the range 8800--9400\,\AA. This indicates that SN 2007gr is a carbon-rich SN~Ic (see a similar argument by V08). The multiple absorption features near 7200\,\AA\ could be identified as Si~I $\lambda\lambda$7020, 7289, 7420. Nevertheless, detailed modeling is needed to confirm these line identifications.

\subsection{Evolution of the Spectra}

Figure 8 shows the spectral comparison of SN 2007gr with some notable SNe~Ic such as SN 1994I (Filippenko et al. 1995), SN 2002ap (Foley et al. 2003), and SN 2004aw (Taubenberger et al. 2006) at several epochs ($\sim 0$ day, $\sim 2$ weeks, $\sim 5$ months, and $\sim 13$ months since $B$-band maximum light). SN 2008D (Modjaz et al. 2009), a SN~Ib, is also plotted for comparison. All of the spectra have been corrected for the redshift of the host galaxy.

\begin{figure}
\begin{center}
\includegraphics[angle=0,width=0.5\textwidth]{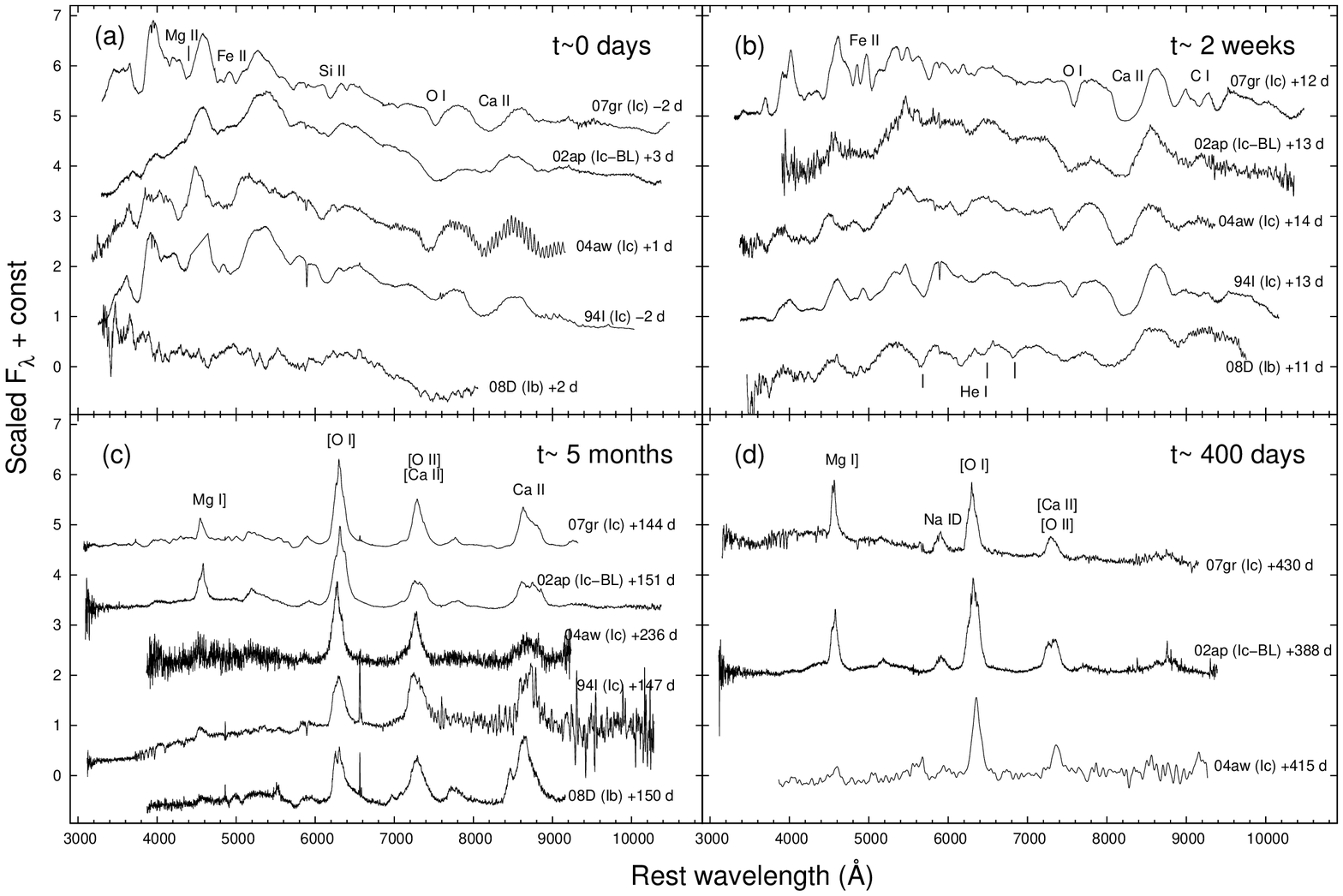}
\end{center}
\caption{\footnotesize
Spectral comparison of SN 2007gr with SNe 1994I (Filippenko et al. 1995), 2002ap (Foley et al. 2003), 2004aw (Taubenberger et al. 2006), and 2008D (Modjaz et al. 2009) at several selected epochs ($t \approx 0$ days, 2 weeks, 5 months, and 400 days).}
\label{fig:sp_com}
\end{figure}

Figure 8a shows the comparison around maximum light. Both SN 2007gr and the comparison SNe reveal features of intermediate-mass elements such as Ca~II, Si~II, Mg~II, and O~I, as well as iron-group elements (Fe~II). Of our SN~Ic sample, SN 2007gr exhibits the narrowest spectral lines, while SN 2002ap shows broader lines with less structure (especially in the region 4000--4500\,\AA). The expansion velocity of SN 2007gr is $\sim 7000$\,km\,s$^{-1}$ at maximum brightness as estimated using the Si~II $\lambda$6355 line, compared to a velocity of $\sim$\,18,000\,km\,s$^{-1}$ at $B$ maximum for SN 2002ap. This suggests that SN 2007gr is less energetic than SN 2002ap because the former has a smaller ejected mass (see discussion in \S 5).

Figure 8b shows the comparison at about two weeks after $B$-band maximum. As in the comparison SNe, the spectrum of SN 2007gr has evolved while maintaining most of its characteristics. The Ca~II NIR features gain strength, while the Si~II lines become weaker and are nearly invisible in the spectrum. One notable aspect of SN 2007gr is the appearance of three narrow absorption features at 4800--5000\,\AA, which may correspond to Fe~II $\lambda\lambda$4924, 5018, 5169 lines. However, these features are less structured in the spectra of the comparison SNe, likely because of line blending. Moreover, SN 2007gr exhibits triple absorption features near 9000\,\AA, probably attributable to C~I lines, which are also visible in the spectrum of SN 1994I. In contrast, only one broad feature is seen at 8800\,\AA\ in spectra of SN 2002ap and SN 2004aw. Such a difference in the absorption-line widths provides additional evidence that SN 2007gr had a lower energy per unit mass.

The comparison at $t \approx 5$ months is shown in Figure 8c. The spectra are now dominated primarily by the emission lines [O~I] $\lambda\lambda$6300, 6364, a blend of [Ca~II] $\lambda\lambda$7291, 7324 and [O~II] $\lambda\lambda$7300, 7330, and the Ca~II NIR triplet. These lines generally have asymmetric structures, with the profiles varying for different compositions and SNe. For example, the [O~I] $\lambda\lambda$6300, 6364 doublet displays a stronger red peak in SNe 1994I, 2002ap, and 2007gr, but seems to have a stronger blue peak in SN 2004aw. While the Mg~I] line exhibits a prominent blue peak in SN 2007gr, it has a strong red component in SN 2002ap. Such diversity in the line profiles is likely related to the asymmetric distribution of the ejecta (Milisavljevic et al. 2010). The profile of the Ca~II NIR triplet is also different for the SNe~Ib/c in our comparison sample. However, it is difficult to use the Ca~II NIR feature to probe the ejecta structure because it can form in different layers of the ejecta owing to a lower ionization energy, and it may also suffer possible contamination from the [C~I] $\lambda$8727 line (Matheson et al. 2000).

A common feature of SN 2007gr and SN 2002ap is the strong Mg~I] $\lambda$4571 emission, with the intensity ratio of Mg~I] to [O~I] being $0.177 \pm 0.049$ and $0.244 \pm 0.045$, respectively. The corresponding ratios are $0.103 \pm 0.084$ and $0.084 \pm 0.095$ in SN 1994I and SN 2004aw, respectively. Differences in the Mg~I]/[O~I] ratio may be interpreted as variations in the degree of stripping of the envelopes experienced by the progenitor star, with a higher ratio implying a greater degree of stripping as more of the inner layers are exposed to the observers (e.g., Foley et al. 2003). This is consistent with the trend that the observed Mg~I]/[O~I] ratio increases with time at later phases. However, a very high ratio of Mg~I]/[O~I] was also reported for the Type IIb SN 2001ig, whose outer layers of H and He were only partially stripped (Silverman et al. 2009), suggesting other possible explanations such as mixing. The higher Mg~I]/[O~I] ratio could be also explained if less of the C/O layer is observed owing to a lower transparency of the ejecta (H09). Moreover, the measured line ratios may be influenced by changes in the environment, as the [O~I] and Mg~I] lines are formed via different mechanisms (Kozma \& Fransson 1998).

A comparison of the very late-time nebular spectra is shown in Figure 8d. The Mg~I] $\lambda$4571 emission becomes noticeably stronger in SN 2007gr and SN 2002ap, with the Mg~I]/[O~I] ratio continuously increasing. However, the Mg~I] line can barely be observed in SN 2004aw, and the increasing Mg~I]/[O~I] trend does not hold for it. By $t \approx 400$ days after maximum light, the Mg~I] $\lambda$ 4571 emission line in SN 2007gr develops a nearly symmetric double-peaked profile, while the line profile in SN 2002ap still maintains the asymmetric structure seen earlier. A common feature for SN 2007gr and the other two comparison SNe is the weakening of the [Ca~II] $\lambda\lambda$7291, 7324 plus [O~II] $\lambda\lambda$7300, 7330 blend and the strengthening of the Na~I lines with time. This can be explained with a gradual decrease in the density of the ejecta (Filippenko et al. 1990).

\subsection{Ejecta Velocities}

The expansion velocities of the ejecta were estimated by fitting a Gaussian profile to the minimum in the absorption lines. The spectra presented by V08 were also included in our analysis. The velocity evolution of different ions is shown in Figure 9. All velocities have been corrected for the redshifts of the host galaxy and the relativistic effect.

\begin{figure}
\begin{center}
\includegraphics[angle=0,width=0.5\textwidth]{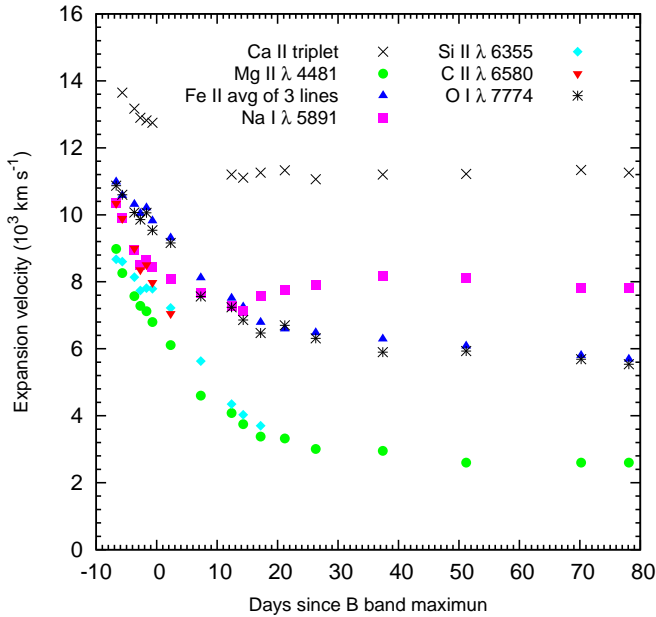}
\end{center}
\caption{\footnotesize
The evolution of the expansion velocities of SN 2007gr measured from the absorption minima of different spectral lines.
}
\label{fig:vol}
\end{figure}

The expansion velocity inferred from the Ca~II NIR triplet is apparently higher than for other elements. It evolved rapidly from $\sim$\,14,000\,km\,s$^{-1}$ at $t = -8$ days to $<$\,11,000\,km\,s$^{-1}$ at $t = +10$ days and maintained a constant velocity thereafter. This indicates that a considerable amount of calcium was distributed in the outermost layers of the ejecta. Such a distribution is perhaps caused by a primordial abundance of calcium that is mixed with the outer part of the envelope (e.g., Li \& McCray 1993) and/or nucleosynthesis from an interaction of SN ejecta with circumstellar material (Gerardy et al. 2004). After $t \approx 20$ days from maximum light, the velocity inferred from other elements such as Na, Mg, O, and Fe also remains nearly constant.

After the $B$-band maximum, Si~II $\lambda$6355 becomes weaker, and blending with Si~I lines makes the measurement of its velocity quite uncertain. In the spectra taken at $t \approx 20$ days the Si features can be hardly detected. The expansion velocity of Fe~II is estimated using the average measured from the three absorption lines at 4924, 5018, and 5169\,\AA; it follows a trend similar to that of O~I $\lambda$7774, decreasing from $\sim$\,11,000\,km\,s$^{-1}$ in the pre-maximum phase to $< 7000$\,km\,s$^{-1}$ at $t > 20$ days. It is worth pointing out that the measured velocity of Na~I increased at $t = 15$ days, possibly because of recombination of Na~II to Na~I (H09).

\section{Discussion}

\subsection{Progenitor of SN 2007gr}

Pre-explosion images of SN 2007gr were obtained on 2001 July 3 with the WFPC2 onboard {\it HST} as part of program GO-9042 (PI S. Smartt). The post-explosion WFPC2 images were obtained about 450 days after maximum light as mentioned in \S 2.2. In Figure 10, we show the pre- and post-explosion {\it HST} WFPC2 images of the site of SN 2007gr.

To accurately determine the explosion site of SN 2007gr, we use a total of 14 stars in the F450W image and 12 stars in the F814W image to establish the geometric transformations from the post-explosion {\it HST} images to the pre-explosion ones. This yields a pixel position of [251, 119] for the SN position in the WF2 pre-explosion image (F450W), with a root-mean-square uncertainty of $\sim 0''.02$. A similar SN position is obtained in the pre-explosion F814W-band image, with an offset $< 0.5$ pixel. As can be seen in Figure 10, no object was detected within the position uncertainty of the SN site in F450W or F814W images. Our result is consistent with that obtained by Crockett et al. (2008) using adaptive-optics observations on the 8.1\,m Gemini North Telescope.

However, there is a bright, point-like source located at a distance of $\sim\,0''.13$ from the SN site, which corresponds to $\sim$\,6.6\,pc if a distance of 10.60\,Mpc is assumed for NGC 1058 (Schmidt et al. 1994). The PSF profile of this object is found to be similar to that of SN 2007gr, implying that it may be a single star or a compact star cluster. DOLPHOT photometry of this source in the pre-explosion WFPC2 images gives F450W $= 21.37 \pm 0.03$ mag and F814W $= 20.82 \pm 0.03$ mag, slightly brighter than the values obtained by Crockett et al. (2008) using HSTphot. We further examined the post-explosion images, yielding $21.47 \pm 0.03$ mag for F450W, $21.29 \pm 0.02$ mag for F555W, $21.12 \pm 0.03$ mag for F675W, and $20.85 \pm 0.02$ mag for F814W. Assuming the photometric result from DOLPHOT is more accurate than that from HSTphot, the star cluster close to SN 2007gr became slightly dimmer in the post-explosion images (especially in F450W), thus favoring the hypothesis that the progenitor of SN 2007gr was a member of the closest star cluster. However, this possible change in magnitude might suffer from light contamination from the supernova, and it should be checked with future {\it HST} images. Adopting the extinction correction and distance to NGC 1058 obtained in \S 4, we find absolute magnitudes of $M_{\rm F450W} = -9.01 \pm 0.35$ mag, $M_{\rm F555W} = -9.12 \pm 0.35$ mag, $M_{\rm F675W} = -9.22 \pm 0.35$ mag, and $M_{\rm F814W} = -9.43 \pm 0.35$ mag for the post-explosion images.

To probe the nature of this object, we fit the observed spectral energy distribution (SED) with distinct theoretical models of a single star and a star cluster. The star spectra are taken from the stellar spectral flux library by Pickles (1998), while the star-cluster spectra are computed with the {\it starburst99} code (Leitherer et al. 1999). Note that the SEDs of clusters or single stars are significantly affected by the adopted metallicity. For SN 2007gr, the relative oxygen abundance log(O/H) + 12 is estimated to be $8.50 \pm 13$ dex at the explosion site according to Modjaz et al. (2011), which is about 37\% of the solar value. We thus adopt a subsolar metallicity of $Z=0.008$ for the SED model in the analysis. A $\chi^{2}$ test is used to examine which members of the SED families are most compatible with the observed SED.

Figure 11 shows the {\it HST} WFPC2 integrated photometry of the object closest to the SN and the best-fit SED models. Fitting the {\it HST} photometry for the stellar model yields the solution of a yellow (F0-type) supergiant, with a $\chi^{2}$ of 8.7 for 4 data points. Besides this source, there are also several others around the SN site with similarly high luminosity (see also Figure 2 of Crockett et al. 2008). This suggests that these brighter sources (including the one closest to the SN) are less likely to be single stars because of their short lifetimes and scarcity in spiral galaxies. On the other hand, the star-cluster model seems to yield a better fit than single stars, with a $\chi ^2$ value of 4.8 for the post-explosion images and 0.9 for the pre-explosion ones.

Both of the pre- and post-explosion SEDs favor a star cluster at an age of 30--38\,Myr, with a turn-off mass of $\sim 8$--9\,M$_{\odot}$ (Girardi et al. 2000; Marigo et al. 2008). Such a low mass would not produce a Wolf-Rayet star via single-star evolution, and it is therefore expected that the progenitor of SN 2007gr was a low-mass Wolf-Rayet star resulting from an interacting binary. We note that Crockett et al. (2008) found two solutions for the age of this assumed cluster: 7.0 Myr and 20--30 Myr. This difference is caused primarily by the solar metallicity used in their fitting (versus $Z = 0.008$ adopted in our analysis). We caution, however, that SN 2007gr does not really coincide with the assumed star cluster, and it is possible that the supernova is not a member of this star cluster. The site should be scrutinized with {\it HST} in several years, when the supernova becomes significantly dimmer.

\begin{figure}
\begin{center}
\includegraphics[angle=0,width=0.5\textwidth]{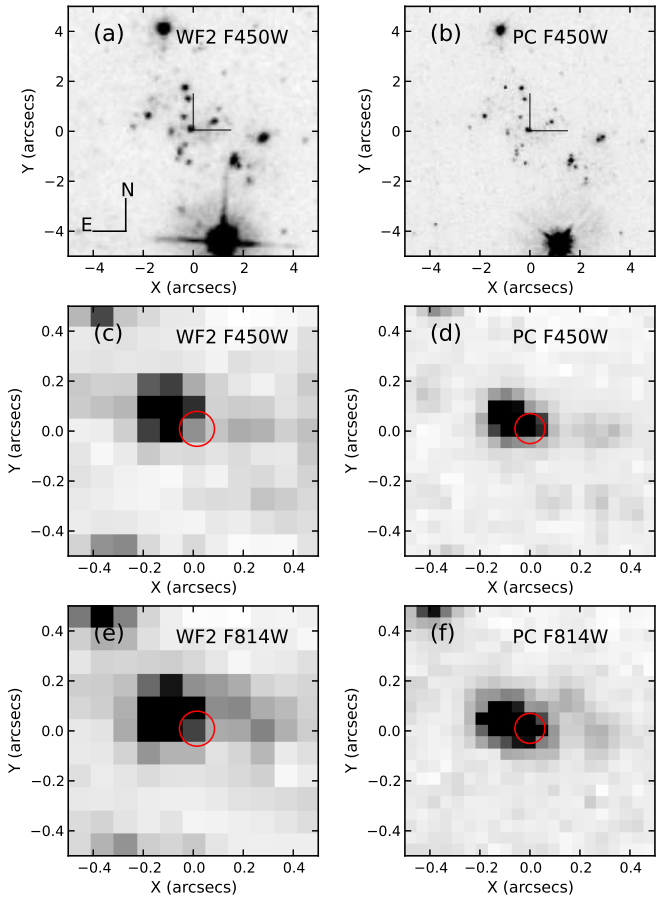}
\end{center}
\caption{\footnotesize
Pre- and post-explosion {\it HST} WFPC2 images of the site of SN 2007gr. Panels (a) and (c) show the pre-explosion WFPC2 F450W image obtained on 2001 July 3; panels (b) and (d) show the post-explosion WFPC2 F450W image obtained on 2008 November 25. Panels (e) and (f) are the pre- and post-explosion F814W images.
}
\label{fig:hst}
\end{figure}

\begin{figure}
\begin{center}
\includegraphics[angle=0,width=0.5\textwidth]{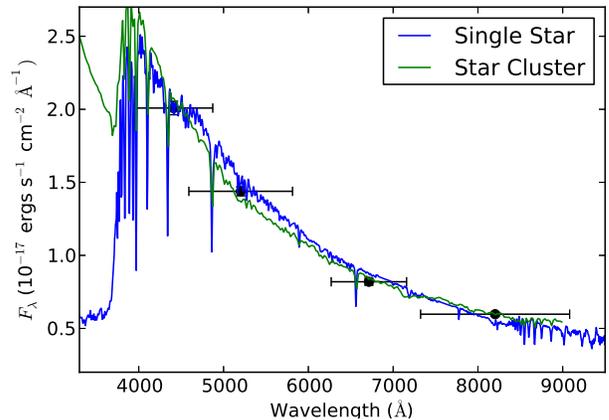}
\end{center}
\caption{\footnotesize
Best-fit SED for the object close to SN 2007gr. See text for detailed of the model parameters. The error bars on the abscissa are the width of the {\it HST} filters.
}
\label{fig:sed}
\end{figure}

\subsection{Asymmetric Line Profiles in Nebular Spectra}

The emission-line profiles in the nebular spectra of SNe are powerful probes of the geometric structure of the ejecta in the explosion. The forbidden lines of oxygen, magnesium, and calcium are the strongest lines in the nebular phase, and have been usually used for studies of this kind. For example, the double-peaked [O~I] $\lambda$6300 profile seen in some SNe~Ib/c has been attributed to a geometric effect of a torus-like structure of O-rich ejecta (Maeda et al. 2008; Modjaz et al. 2008). Taubenberger et al. (2009) compared the line profiles of Mg~I] $\lambda$4571 and [O~I] $\lambda$6300 with a sample of 39 core-collapse SNe, suggesting that O and Mg are similarly distributed within the ejecta. Milisavljevic et al. (2010) recently questioned the torus-like geometry of the ejecta as an explanation of the observed double-peaked [O~I] $\lambda\lambda$6300, 6364 emission through a reexamination of the line profiles. In our analysis, we compare the profiles of Mg~I] and [O~I] in the six spectra of SN 2007gr obtained from $t=+78$ days through +430 days (Fig. 10), with an attempt to put some constraints on the ejecta geometry of SN 2007gr.

At $t=+78$ days, [O~I] displayed a nearly symmetric double-peaked structure, with the peaks centered at 6260\,\AA\ and 6300\,\AA. The feature became asymmetric in the $t=+129$ day spectrum and thereafter, but the gradually decreasing intensity ratio of the blue peak to the red peak is inconsistent with the expected evolution of the intensity ratio of [O~I] $\lambda$6300 and $\lambda$6364. Moreover, we notice that the wavelength difference between these two peaks is 40\,\AA, smaller than the true separation of the [O~I] $\lambda\lambda$6300, 6364 doublet. These results indicate that the emission peaks seen at 6300\,\AA\ and 6260\,\AA\ are unlikely to be from the two distinct components of the [O I] $\lambda\lambda$6300, 6364 doublet. Attributing the blue peak seen at 6260\,\AA\ to blueshifted [O~I] $\lambda$6300 yields a velocity shift of about 1800\,km\,s$^{-1}$. As can be seen in Figure 12a, the multiple emission peaks do not show any noticeable shift with time in velocity space. Moreover, the relative strength of the blueshifted peak (marked by the dotted lines) decreases gradually relative to the emission peak near zero velocity (marked by the dashed lines).

Close inspection of Figure 12a reveals that there are two other minor emission features with central wavelengths at 6320\,\AA\ and 6360\,\AA. We suggest that these are likely produced by blueshifted and rest-frame [O~I] $\lambda$6364, as the velocity shift inferred from the blue peak is $\sim 1800$\,km\,s$^{-1}$, consistent with the value derived from the [O~I] $\lambda$6300 line. To confirm our identifications, we examine the velocity structure of the Mg~I] $\lambda$4571 line in Figure 12b. As can be seen, the Mg~I] $\lambda$4571 line also displays a double-peaked feature in SN 2007gr, with the blue-side peak being shifted by about $\sim 1800$\,km\,s$^{-1}$ with respect to the rest-frame (or the red-side) component. Figure 12c shows the comparison of the [O~I] $\lambda\lambda$6300, 6364 doublet with the scaled Mg~I] $\lambda$4571 line profile in the $t=+430$ day spectrum.

The Mg~I] line shows a stronger blueshifted peak in the early nebular phase, and the rest-frame peak appears in the later spectra (after $t = 144$ days). In comparison, the [O~I] $\lambda$6300 line exhibits both the rest-frame and blueshifted components during the nebular phase, and the blueshifted peak gradually weakens with time. This feature can be understood in light of torus-like ejecta produced by an aspherical explosion. The blueshifted peak, which arises in the ejecta closer to the observer, is present in all of the nebular spectra. The emission from the rear side of the SN is partially scattered and/or absorbed by the ejecta, and it gains strength at late phases when the ejecta become optically thin.

We note that the flux ratio of the blueshifted to the rest-frame peaks is different for the Mg~I] and [O~I] lines. The strong blueshifted peak in the Mg~I] line leads to the mismatch between the [O~I] and Mg~I] profiles in the early nebular spectra (see also the mismatch shown by H09). Even in the $t=+430$ day spectrum, the blueshifted peak of the Mg~I] line is slightly stronger than that of the [O~I] line (see Fig. 12c). This difference suggests that oxygen and magnesium have different distributions within the ejecta. More magnesium is located in the outer part, while most of the oxygen is located on the rear side of the ejecta.

At $t=+430$ days, the flux ratio of [O~I] $\lambda$6300 to [O~I] $\lambda$6364 \AA\ is estimated to be 3:1. According to theoretical calculations (Spyromilio 1991; Williams 1994), this ratio is about 1:1 when the ejecta are optically thick at early times. Examining the spectra at $t = 129$ days, we find that the flux ratio of the [O~I] doublet is also close to 3:1. This suggests that SN 2007gr already became optically thin in the early nebular phase, and hence the explosion ejected a small mass of material. The small ejecta mass, together with the low observed expansion velocities (hence low kinetic energy per unit mass), imply a low-energy explosion for SN 2007gr.

\begin{figure}
\begin{center}
\includegraphics[angle=0,width=0.5\textwidth]{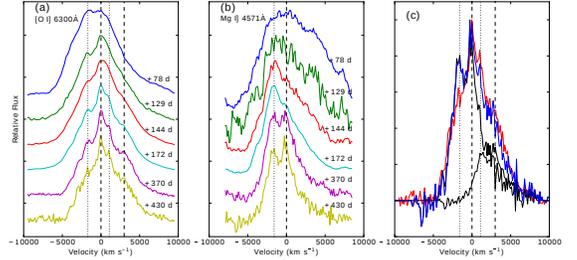}
\end{center}
\caption{\footnotesize
Comparison of line profiles of [O~I] $\lambda$6300 and Mg~I] $\lambda$4571 in SN 2007gr at six phases. In panels (a) and (b), the dashed lines mark the rest wavelengths 6300, 6364, and 4571\,\AA, and the dotted lines mark the blueshifted peaks at velocities of about 1800\,km\,s$^{-1}$. Panel (c) shows a comparison of the modified Mg~I] (blue) and the [O~I] (red) line profiles in the $t=+430$ day spectrum.
The Mg~I] profile includes an artificial component with a velocity offset of 3000\,km\,s$^{-1}$ (equivalent to the 64\,\AA\ separation of the two [O~I] lines); the intensity of the red-side profile was scaled by 1/3, the expected intensity ration of [O~I] $\lambda$6364 to $\lambda$6300. These two Mg~I] components are marked by the black lines.
}
\label{fig:neb_c}
\end{figure}

\section{Conclusions}

In this paper we present extensive optical photometry and spectra of the Type Ic SN 2007gr. This object shows a remarkable resemblance to SN 2002ap in terms of the light and color curves. However, SN 2007gr has narrower line features than the latter and other normal SNe~Ib/c.

The optical spectra, covering from $t \approx -8$ days to +430 days from maximum light, provide significant constraints on the physical properties of SN 2007gr. By modeling the $t \approx -8$ day spectrum with SYNOW, we identify the absorption feature at 5500\,\AA\ as possibly being He~I $\lambda$5876, corresponding to a helium layer located at very high velocity ($\sim$\,19,000\,km\,s$^{-1}$). At $t=+12$ days from maximum light, SN 2007gr shows prominent narrower C~I lines around 9000\,\AA, which are usually very weak in SNe~Ic, indicating that the outer layers of SN 2007gr are abundant in carbon. Based on the possible detection of helium lines in the very early-time spectrum and the pronounced carbon lines in the later spectra, we suggest that SN 2007gr represents a transition object between SNe~Ib and SNe~Ic. The progenitor thus lost its hydrogen envelope but may have retained a small amount of helium in the outer layer before the explosion.

According to estimates of the age of the star cluster in which SN 2007gr may be present, there are two solutions for the turn-off mass and hence the initial mass of the progenitor star: $\sim 30$\,M$_{\odot}$ and $\sim 10$\,M$_{\odot}$ (Crockett et al. 2008). Our results presented in this paper seem to be more consistent with the latter solution, with a lower-mass Wolf-Rayet carbon (WC) star in a close binary (see Crowther 2007), according to the following arguments: (1) the narrower lines seen in the spectra of SN 2007gr favor a relatively small amount of energy per unit of ejecta mass; (2) the mass of synthesized $^{56}$Ni and the ejecta mass derived from the quasi-bolometric light curve around maximum light observed in SN 2007gr are smaller than the corresponding values predicted by the single massive WC star model ($\sim 30$\,M$_{\odot}$; Smartt 2009); (3) the intensity ratio of 3:1 for [O~I] $\lambda$6300 to [O~I] $\lambda$6364 seen in the $t=129$ day spectrum indicates that the ejecta are already optically thin in the early nebular phase, suggestive of a smaller mass ejected during the explosion; and (4) the pre- and post-explosion {\it HST} observations suggest that the SN is likely within a compact star cluster having a turn-off mass of $\sim 8$--9\,M$_{\odot}$.

We compare the evolution of the double-peaked structure of the Mg~I] and [O~I] lines in the nebular spectra. Based on the velocity structure of the Mg~I] and [O~I] lines, we obtain a more secure identification of the blueshifted and rest-frame components of the [O~I] $\lambda\lambda$6300, 6364 doublet. Moreover, the similarity of the Mg and O elements in velocity space and the evolution of the flux ratio of their blueshifted and rest-frame components indicate that they have different distributions in the ejecta. Thus, the asymmetric double-peaked profiles seen in [O~I] and Mg~I] could be produced by an aspherical explosion whose ejecta have a torus-like structure (Maeda et al. 2002).

\acknowledgments

We are grateful to the anonymous referee for his/her constructive suggestions which helped improve the paper. We thank the NAOC, Lick, and Keck Observatory staffs, as well as Maryam Modjaz and Ryan Foley, for their assistance with the observations. Some of the data presented herein were obtained at the W. M. Keck Observatory, which is operated as a scientific partnership among the California Institute of Technology, the University of California, and NASA; the observatory was made possible by the generous financial support of the W. M. Keck Foundation. This work is supported by the Major State Basic Research Development Program (2013CB834903), National Natural Science Foundation of China (NSFC grants 11073013, 11178003, 11325313), and the Foundation of Tsinghua University (2011Z02170). A.V.F.'s group at UC Berkeley is grateful for financial assistance from NSF grant AST-1211916, the TABASGO Foundation, and the Christopher R. Redlich Fund. This work was also supported by NASA through grants GO-10877 and AR-12623 from the Space Telescope Science Institute, which is operated by the Associated Universities for Research in Astronomy, Inc., under NASA contract NAS 5-26555. J. M. Silverman is supported by an NSF Astronomy and Astrophysics Postdoctoral Fellowship under award AST-1302771.

\begin{deluxetable}{cccccc}
\tabletypesize{\scriptsize}
\tablecaption{Instrumental Color Terms for Different Telescopes.\tablenotemark{a}}
\tablewidth{0pt}
\tablehead{
\colhead{Telescope} & \colhead{$U$ (mag)} & \colhead{$B$ (mag)} & \colhead{$V$ (mag)} &
\colhead{$R$ (mag)} & \colhead{$I$ (mag)}}
\startdata
KAIT4\tablenotemark{b} & $-$0.085(0.017) &$-$0.043(0.011) & 0.035(0.007) & 0.070(0.012) & $-$0.010(0.006) \\
Nickel & $-$0.080(0.010) &$-$0.080(0.011) & 0.060(0.007) & 0.100(0.012) & $-$0.035(0.006) \\
TNT  & $-$0.124(0.016) &$-$0.132(0.004) & 0.080(0.004) & 0.106(0.006) & $-$0.037(0.003) \\
\enddata
\tablenotetext{a}{$1\sigma$ uncertainties are given in parentheses.}
\tablenotetext{b}{KAIT4 stands for the fourth filter set used at KAIT; see Ganeshalingam et al. (2010) for details.}
\end{deluxetable}

\begin{deluxetable}{cccccc}
\tabletypesize{\scriptsize}
\tablecaption{Photometric Standard Stars in the SN 2007gr Field.\tablenotemark{a}. \label{tab:CS} }
\tablewidth{0pt}
\tablehead{
\colhead{Star} & \colhead{$U$ (mag)} & \colhead{$B$ (mag)} & \colhead{$V$ (mag)} &
\colhead{$R$ (mag)} & \colhead{$I$ (mag)}
}
\startdata
1 & 14.068(0.005) & 13.846(0.003) & 13.334(0.002) & 12.980(0.003) & 12.693(0.003) \\
2 & 15.951(0.011) & 15.525(0.006) & 14.698(0.003) & 14.212(0.004) & 13.767(0.004) \\
3 & 15.137(0.007) & 15.058(0.004) & 14.523(0.002) & 14.198(0.004) & 13.881(0.004) \\
4 & 15.843(0.011) & 15.506(0.006) & 14.788(0.003) & 14.374(0.004) & 13.997(0.005) \\
5 & 12.473(0.002) & 12.294(0.001) & 11.979(0.001) & 11.801(0.001) & 11.627(0.001) \\
6 & 15.640(0.010) & 15.429(0.006) & 14.795(0.003) & 14.422(0.005) & 14.082(0.005) \\
7 & 14.314(0.004) & 14.239(0.003) & 13.678(0.001) & 13.336(0.002) & 13.012(0.002) \\
8 & 15.353(0.008) & 15.206(0.005) & 14.547(0.002) & 14.137(0.004) & 13.750(0.004) \\
9 & 14.037(0.003) & 13.868(0.002) & 13.250(0.001) & 12.851(0.002) & 12.488(0.002) \\
10 & 16.647(0.018) & 15.540(0.005) & 14.436(0.002) & 13.845(0.003) & 13.323(0.003) \\
11 & 14.155(0.004) & 14.052(0.002) & 13.463(0.001) & 13.117(0.002) & 12.791(0.002) \\
12 & 15.811(0.010) & 15.112(0.004) & 14.220(0.002) & 13.654(0.003) & 13.181(0.003) \\
\enddata
\tablenotetext{a}{See Figure~1 for a chart of SN 2007gr and the comparison stars. $1\sigma$ uncertainties are given in parentheses.}
\end{deluxetable}

\begin{deluxetable}{ccccccccc}
\tabletypesize{\tiny}
\tablecaption{S-Corrected Optical Photometry of SN 2007gr.\tablenotemark{a} \label{tab:LC}}
\tablewidth{0pt}
\tablehead{
\colhead{Ut Date} & \colhead{JD$-$2,450,000} & \colhead{Phase\tablenotemark{b}} & \colhead{$U$ (mag)} & \colhead{$B$ (mag)} & \colhead{$V$ (mag)} & \colhead{$R$ (mag)} & \colhead{$I$ (mag)} & \colhead{Instrument}
}
\startdata
2007 Aug 15 & 4328.43 & -8.32 & \nodata & 14.297 (0.030) & 13.995 (0.020) & 13.848 (0.021) & 13.750 (0.030) & KAIT4\tablenotemark{c} \\
2007 Aug 16 & 4329.45 & -7.30 & \nodata & 14.077 (0.030) & 13.726 (0.020) & 13.580 (0.020) & 13.544 (0.030) & KAIT4 \\
2007 Aug 17 & 4330.31 & -6.44 & 13.708 (0.036) & 14.061 (0.023) & 13.668 (0.023) & 13.525 (0.142) & 13.438 (0.023) & TNT \\
2007 Aug 18 & 4331.49 & -5.26 & \nodata & 13.692 (0.030) & 13.366 (0.020) & 13.261 (0.020) & 13.187 (0.030) & KAIT4 \\
2007 Aug 19 & 4332.45 & -4.30 & \nodata & 13.595 (0.030) & 13.264 (0.020) & 13.124 (0.021) & 13.025 (0.030) & Nickel \\
2007 Aug 19 & 4332.46 & -4.29 & \nodata & 13.644 (0.030) & 13.246 (0.020) & 13.142 (0.020) & 13.081 (0.030) & KAIT4 \\
2007 Aug 20 & 4333.42 & -3.33 & \nodata & 13.503 (0.030) & 13.140 (0.020) & 13.042 (0.020) & 12.913 (0.030) & Nickel \\
2007 Aug 20 & 4333.45 & -3.30 & \nodata & 13.520 (0.030) & 13.196 (0.020) & 13.055 (0.020) & 12.989 (0.030) & KAIT4 \\
2007 Aug 21 & 4334.35 & -2.40 & 13.377 (0.105) & 13.571 (0.026) & 13.133 (0.023) & 13.024 (0.020) & 12.870 (0.019) & TNT \\
2007 Aug 21 & 4334.45 & -2.30 & \nodata & 13.496 (0.030) & 13.034 (0.020) & 12.933 (0.020) & 12.871 (0.030) & KAIT4 \\
2007 Aug 22 & 4335.48 & -1.27 & \nodata & 13.453 (0.030) & 12.941 (0.020) & 12.885 (0.020) & 12.803 (0.030) & KAIT4 \\
2007 Aug 23 & 4336.49 & -0.26 & \nodata & 13.468 (0.030) & 12.973 (0.020) & 12.828 (0.020) & 12.753 (0.030) & KAIT4 \\
2007 Aug 24 & 4337.35 & 0.70 & \nodata & 13.508 (0.030) & 12.981 (0.020) & 12.834 (0.020) & 12.716 (0.030) & KAIT4 \\
2007 Aug 26 & 4339.45 & 2.70 & \nodata & 13.532 (0.030) & 12.930 (0.021) & 12.748 (0.020) & 12.663 (0.030) & KAIT4 \\
2007 Aug 27 & 4340.31 & 3.56 & 13.754 (0.082) & 13.603 (0.020) & 12.942 (0.022) & 12.752 (0.027) & 12.601 (0.029) & TNT \\
2007 Aug 27 & 4340.35 & 3.60 & \nodata & 13.547 (0.030) & 12.939 (0.020) & 12.734 (0.020) & 12.586 (0.030) & Nickel \\
2007 Aug 27 & 4340.40 & 3.65 & \nodata & \nodata & 12.950 (0.021) & 12.748 (0.021) & 12.659 (0.043) & KAIT4 \\
2007 Aug 28 & 4341.48 & 4.73 & \nodata & 13.603 (0.030) & 12.926 (0.020) & 12.697 (0.021) & 12.633 (0.030) & KAIT4 \\
2007 Aug 29 & 4342.44 & 5.69 & \nodata & 13.755 (0.030) & 12.979 (0.020) & 12.801 (0.023) & 12.613 (0.030) & KAIT4 \\
2007 Aug 30 & 4343.25 & 6.50 & 14.088 (0.272) & 13.777 (0.023) & 13.006 (0.019) & 12.812 (0.026) & 12.590 (0.025) & TNT \\
2007 Aug 30 & 4343.44 & 6.69 & \nodata & \nodata & 13.009 (0.020) & 12.742 (0.021) & \nodata & KAIT4 \\
2007 Aug 31 & 4344.33 & 7.58 & 14.644 (0.379) & 13.901 (0.026) & 13.075 (0.021) & 12.803 (0.025) & 12.632 (0.026) & TNT \\
2007 Aug 31 & 4344.44 & 7.69 & \nodata & 13.934 (0.030) & 13.042 (0.020) & 12.788 (0.022) & 12.674 (0.030) & KAIT4 \\
2007 Aug 31 & 4344.46 & 7.71 & \nodata & 13.895 (0.030) & 13.076 (0.020) & 12.800 (0.020) & 12.616 (0.030) & Nickel \\
2007 Sep 2 & 4346.50 & 9.75 & \nodata & 14.183 (0.032) & 13.173 (0.023) & 12.922 (0.020) & 12.737 (0.031) & KAIT4 \\
2007 Sep 3 & 4347.24 & 10.49 & 14.775 (0.042) & 14.186 (0.034) & 13.201 (0.042) & 12.896 (0.088) & 12.747 (0.237) & TNT \\
2007 Sep 3 & 4347.41 & 10.66 & \nodata & 14.250 (0.031) & 13.290 (0.020) & 12.961 (0.021) & 12.761 (0.030) & KAIT4 \\
2007 Sep 4 & 4348.36 & 11.61 & 14.961 (0.037) & 14.348 (0.026) & 13.324 (0.018) & 13.020 (0.026) & 12.758 (0.012) & TNT \\
2007 Sep 4 & 4348.46 & 11.71 & \nodata & 14.383 (0.030) & 13.368 (0.020) & 13.010 (0.020) & 12.805 (0.030) & KAIT4 \\
2007 Sep 5 & 4349.35 & 12.60 & 15.147 (0.042) & 14.441 (0.028) & 13.392 (0.028) & 13.021 (0.021) & 12.735 (0.278) & TNT \\
2007 Sep 5 & 4349.47 & 12.72 & \nodata & 14.515 (0.030) & 13.444 (0.020) & 13.059 (0.020) & 12.818 (0.030) & KAIT4 \\
2007 Sep 6 & 4350.29 & 13.54 & 15.346 (0.059) & 14.619 (0.027) & 13.517 (0.028) & 13.121 (0.021) & 12.849 (0.017) & TNT \\
2007 Sep 7 & 4351.29 & 14.54 & \nodata & 14.697 (0.025) & 13.596 (0.020) & 13.155 (0.026) & 12.863 (0.020) & TNT \\
2007 Sep 8 & 4352.27 & 15.52 & \nodata & 14.809 (0.025) & 13.681 (0.020) & 13.202 (0.020) & 12.914 (0.019) & TNT \\
2007 Sep 9 & 4353.17 & 16.42 & 15.668 (0.173) & 14.905 (0.025) & 13.738 (0.020) & 13.267 (0.018) & 12.944 (0.026) & TNT \\
2007 Sep 9 & 4353.52 & 16.77 & \nodata & 15.003 (0.030) & 13.809 (0.020) & 13.298 (0.027) & 13.021 (0.030) & KAIT4 \\
2007 Sep 10 & 4354.17 & 17.42 & \nodata & 14.951 (0.025) & 13.819 (0.025) & 13.339 (0.021) & 12.981 (0.024) & TNT \\
2007 Sep 11 & 4355.17 & 18.42 & 15.864 (0.061) & 15.043 (0.028) & 13.904 (0.019) & 13.378 (0.019) & 13.038 (0.023) & TNT \\
2007 Sep 11 & 4355.45 & 18.70 & \nodata & 15.148 (0.030) & 13.957 (0.020) & 13.423 (0.020) & 13.106 (0.030) & KAIT4 \\
2007 Sep 13 & 4357.51 & 20.76 & \nodata & 15.354 (0.030) & 14.160 (0.020) & 13.619 (0.020) & 13.235 (0.030) & KAIT4 \\
2007 Sep 14 & 4358.43 & 21.68 & \nodata & 15.430 (0.030) & 14.234 (0.020) & 13.668 (0.020) & 13.269 (0.031) & KAIT4 \\
2007 Sep 14 & 4358.51 & 21.76 & \nodata & 15.402 (0.030) & 14.212 (0.020) & 13.660 (0.020) & 13.275 (0.030) & Nickel \\
2007 Sep 15 & 4359.27 & 22.52 & \nodata & 15.446 (0.025) & 14.250 (0.021) & 13.679 (0.024) & 13.251 (0.025) & TNT \\
2007 Sep 15 & 4359.47 & 22.72 & \nodata & 15.495 (0.030) & 14.281 (0.020) & 13.731 (0.020) & 13.313 (0.030) & KAIT4 \\
2007 Sep 16 & 4360.28 & 23.53 & \nodata & 15.511 (0.024) & 14.326 (0.020) & 13.752 (0.026) & 13.288 (0.021) & TNT \\
2007 Sep 16 & 4360.41 & 23.66 & \nodata & 15.549 (0.030) & 14.370 (0.020) & 13.783 (0.020) & 13.324 (0.030) & KAIT4 \\
2007 Sep 17 & 4361.37 & 24.62 & \nodata & 15.633 (0.031) & 14.432 (0.021) & 13.887 (0.020) & 13.439 (0.030) & KAIT4 \\
2007 Sep 18 & 4362.37 & 25.62 & \nodata & 15.641 (0.031) & 14.460 (0.020) & 13.912 (0.021) & 13.454 (0.030) & KAIT4 \\
2007 Sep 20 & 4364.30 & 27.55 & 16.430 (0.042) & 15.727 (0.027) & 14.563 (0.026) & 13.988 (0.024) & 13.506 (0.028) & TNT \\
2007 Sep 23 & 4367.39 & 30.64 & \nodata & 15.798 (0.030) & 14.677 (0.020) & 14.151 (0.020) & 13.600 (0.030) & KAIT4 \\
2007 Sep 24 & 4367.44 & 30.69 & \nodata & 15.772 (0.030) & 14.740 (0.020) & 14.006 (0.020) & \nodata & Nickel \\
2007 Sep 25 & 4369.31 & 32.56 & \nodata & 15.854 (0.030) & 14.778 (0.020) & 14.226 (0.020) & 13.678 (0.030) & KAIT4 \\
2007 Oct 2 & 4376.36 & 39.61 & \nodata & 15.966 (0.033) & 14.942 (0.020) & 14.405 (0.021) & 13.830 (0.030) & KAIT4 \\
2007 Oct 3 & 4377.38 & 40.63 & \nodata & 16.076 (0.040) & 14.992 (0.043) & 14.383 (0.063) & 13.855 (0.033) & TNT \\
2007 Oct 5 & 4379.37 & 42.62 & \nodata & 16.012 (0.030) & 15.001 (0.020) & 14.507 (0.020) & 13.890 (0.030) & KAIT4 \\
2007 Oct 5 & 4379.48 & 42.73 & \nodata & 15.929 (0.030) & 14.985 (0.020) & 14.459 (0.020) & 13.885 (0.030) & Nickel \\
2007 Oct 6 & 4380.51 & 43.76 & \nodata & 15.976 (0.030) & 14.980 (0.020) & 14.466 (0.020) & 13.892 (0.030) & Nickel \\
2007 Oct 7 & 4381.40 & 44.65 & \nodata & 16.015 (0.031) & 15.013 (0.020) & 14.522 (0.020) & 13.901 (0.030) & KAIT4 \\
2007 Oct 10 & 4384.31 & 47.56 & \nodata & 16.013 (0.031) & 15.066 (0.020) & 14.603 (0.020) & 13.974 (0.030) & KAIT4 \\
2007 Oct 13 & 4387.39 & 50.64 & \nodata & 16.063 (0.032) & 15.099 (0.020) & 14.636 (0.027) & 13.990 (0.030) & KAIT4 \\
2007 Oct 16 & 4390.24 & 53.49 & 16.523 (0.030) & 16.100 (0.019) & 15.164 (0.023) & 14.677 (0.026) & 14.066 (0.028) & TNT \\
2007 Oct 17 & 4391.39 & 54.64 & \nodata & 16.129 (0.031) & 15.178 (0.020) & 14.755 (0.020) & 14.093 (0.030) & KAIT4 \\
2007 Oct 17 & 4391.49 & 54.74 & \nodata & \nodata & \nodata & 14.745 (0.020) & \nodata & Nickel \\
2007 Oct 18 & 4392.26 & 55.51 & \nodata & 16.088 (0.019) & 15.110 (0.026) & 14.703 (0.026) & 14.080 (0.025) & TNT \\
2007 Oct 20 & 4394.20 & 57.45 & 16.561 (0.078) & 16.121 (0.020) & 15.226 (0.017) & 14.732 (0.025) & 14.110 (0.023) & TNT \\
2007 Oct 21 & 4395.22 & 58.47 & 16.663 (0.041) & 16.150 (0.021) & 15.223 (0.025) & 14.777 (0.024) & 14.129 (0.026) & TNT \\
2007 Oct 21 & 4395.26 & 58.51 & \nodata & 16.141 (0.032) & 15.236 (0.024) & 14.822 (0.020) & 14.126 (0.030) & KAIT4 \\
2007 Oct 22 & 4396.25 & 59.50 & \nodata & 16.165 (0.020) & 15.202 (0.024) & \nodata & 14.150 (0.023) & TNT \\
2007 Oct 24 & 4398.25 & 61.50 & 16.638 (0.145) & 16.164 (0.026) & 15.279 (0.023) & \nodata & 14.262 (0.096) & TNT \\
2007 Oct 27 & 4401.43 & 64.68 & \nodata & \nodata & \nodata & 14.998 (0.021) & \nodata & Nickel \\
2007 Oct 28 & 4402.29 & 65.54 & \nodata & 16.192 (0.033) & 15.331 (0.021) & 14.934 (0.021) & 14.228 (0.030) & KAIT4 \\
2007 Oct 30 & 4404.20 & 67.45 & 16.686 (0.041) & 16.225 (0.022) & 15.373 (0.017) & 14.898 (0.025) & 14.282 (0.019) & TNT \\
2007 Oct 31 & 4405.20 & 68.45 & 16.662 (0.202) & 16.196 (0.026) & 15.375 (0.024) & 14.962 (0.034) & 14.293 (0.020) & TNT \\
2007 Nov 1 & 4406.24 & 69.49 & 16.694 (0.041) & 16.213 (0.020) & 15.479 (0.025) & 14.952 (0.023) & 14.325 (0.022) & TNT \\
2007 Nov 1 & 4406.36 & 69.61 & \nodata & 16.269 (0.030) & 15.464 (0.021) & 15.008 (0.020) & 14.333 (0.030) & KAIT4 \\
2007 Nov 5 & 4410.32 & 73.57 & \nodata & 16.349 (0.031) & 15.478 (0.020) & 15.075 (0.025) & 14.397 (0.030) & KAIT4 \\
2007 Nov 9 & 4414.25 & 77.50 & 16.801 (0.040) & 16.391 (0.018) & 15.491 (0.026) & 15.119 (0.022) & 14.458 (0.023) & TNT \\
2007 Nov 9 & 4414.37 & 77.62 & \nodata & 16.419 (0.030) & 15.577 (0.020) & 15.138 (0.021) & 14.452 (0.030) & KAIT4 \\
2007 Nov 9 & 4414.45 & 77.70 & \nodata & 16.356 (0.030) & 15.576 (0.020) & 15.148 (0.020) & 14.488 (0.030) & Nickel \\
2007 Nov 11 & 4416.16 & 79.41 & \nodata & 16.498 (0.020) & 15.676 (0.019) & 15.201 (0.019) & 14.525 (0.025) & TNT \\
2007 Nov 13 & 4418.35 & 81.60 & \nodata & 16.419 (0.031) & 15.625 (0.022) & 15.217 (0.020) & 14.508 (0.030) & KAIT4 \\
2007 Nov 15 & 4420.22 & 83.47 & 16.905 (0.048) & 16.421 (0.029) & 15.655 (0.018) & 15.231 (0.020) & 14.603 (0.029) & TNT \\
2007 Nov 17 & 4422.34 & 85.59 & \nodata & 16.500 (0.033) & 15.703 (0.020) & 15.303 (0.020) & 14.619 (0.030) & KAIT4 \\
2007 Nov 17 & 4422.43 & 85.68 & \nodata & \nodata & 15.692 (0.020) & 15.290 (0.022) & 14.623 (0.030) & Nickel \\
2007 Nov 20 & 4425.03 & 88.28 & 17.101 (0.053) & 16.584 (0.026) & 15.730 (0.020) & 15.267 (0.022) & 14.649 (0.028) & TNT \\
2007 Nov 25 & 4430.33 & 93.58 & \nodata & 16.581 (0.030) & \nodata & 15.343 (0.020) & 14.744 (0.030) & Nickel \\
2007 Nov 26 & 4431.10 & 94.35 & 17.201 (0.072) & 16.663 (0.015) & 15.820 (0.026) & 15.385 (0.024) & 14.805 (0.020) & TNT \\
2007 Nov 26 & 4431.16 & 94.41 & \nodata & 16.604 (0.030) & 15.882 (0.020) & 15.376 (0.020) & 14.783 (0.030) & Nickel \\
2007 Nov 26 & 4431.19 & 94.44 & \nodata & 16.570 (0.037) & 15.884 (0.020) & 15.351 (0.023) & 14.724 (0.030) & KAIT4 \\
2007 Nov 27 & 4432.27 & 95.52 & \nodata & 16.657 (0.031) & 15.853 (0.034) & 15.422 (0.026) & 14.783 (0.033) & KAIT4 \\
2007 Dec 4 & 4439.31 & 102.56 & \nodata & 16.763 (0.030) & 15.989 (0.020) & 15.499 (0.020) & 14.912 (0.030) & KAIT4 \\
2007 Dec 13 & 4448.22 & 111.47 & \nodata & 16.886 (0.030) & 16.173 (0.023) & 15.648 (0.022) & 15.030 (0.030) & KAIT4 \\
2007 Dec 14 & 4449.32 & 112.57 & \nodata & 16.951 (0.030) & \nodata & 15.661 (0.023) & 15.165 (0.033) & Nickel \\
2007 Dec 22 & 4457.27 & 120.52 & \nodata & 17.027 (0.083) & 16.273 (0.051) & 15.732 (0.032) & 15.244 (0.032) & KAIT4 \\
2007 Dec 24 & 4459.25 & 122.50 & 17.648 (0.178) & 17.013 (0.039) & 16.394 (0.083) & 15.775 (0.116) & \nodata & TNT \\
2007 Dec 30 & 4465.23 & 128.48 & \nodata & 17.187 (0.030) & 16.453 (0.030) & 15.887 (0.031) & 15.352 (0.037) & KAIT4 \\
2008 Jan 1 & 4467.01 & 130.26 & 17.677 (0.130) & 17.276 (0.027) & 16.408 (0.038) & 15.831 (0.024) & 15.427 (0.020) & TNT \\
2008 Jan 6 & 4472.10 & 135.35 & 17.915 (0.118) & 17.328 (0.028) & 16.538 (0.032) & 16.031 (0.023) & 15.442 (0.028) & TNT \\
2008 Jan 7 & 4473.17 & 136.42 & 17.984 (0.147) & 17.348 (0.022) & 16.726 (0.046) & 15.921 (0.024) & 15.578 (0.033) & TNT \\
2008 Jan 8 & 4474.14 & 137.39 & 18.009 (0.197) & 17.395 (0.039) & 16.684 (0.040) & 16.079 (0.043) & 15.569 (0.027) & TNT \\
2008 Jan 9 & 4475.13 & 138.38 & 18.032 (0.126) & 17.398 (0.023) & 16.773 (0.039) & 15.976 (0.132) & 15.682 (0.049) & TNT \\
2008 Jan 11 & 4477.30 & 140.55 & \nodata & 17.373 (0.030) & 16.730 (0.020) & 16.052 (0.021) & 15.603 (0.030) & Nickel \\
2008 Jan 12 & 4478.27 & 141.52 & \nodata & 17.436 (0.030) & 16.747 (0.020) & \nodata & \nodata & Nickel \\
2008 Jan 14 & 4480.23 & 143.48 & \nodata & 17.433 (0.092) & 16.744 (0.040) & 16.103 (0.037) & 15.643 (0.033) & KAIT4 \\
2008 Jan 18 & 4484.22 & 147.47 & \nodata & 17.562 (0.050) & 16.878 (0.023) & 16.155 (0.025) & 15.733 (0.033) & KAIT4 \\
2008 Feb 4 & 4501.11 & 164.36 & \nodata & 17.822 (0.041) & 17.152 (0.036) & 16.379 (0.027) & 16.101 (0.033) & KAIT4 \\
2008 Feb 8 & 4505.18 & 168.43 & \nodata & 17.882 (0.066) & 17.257 (0.028) & 16.461 (0.047) & 16.167 (0.030) & KAIT4 \\
2008 Feb 8 & 4505.21 & 168.46 & \nodata & \nodata & 17.205 (0.020) & 16.484 (0.022) & 16.103 (0.040) & Nickel \\
2008 Feb 12 & 4509.20 & 172.45 & \nodata & 18.031 (0.030) & 17.399 (0.020) & 16.617 (0.023) & 16.194 (0.035) & Nickel \\
2008 Feb 15 & 4512.12 & 175.37 & \nodata & 18.006 (0.071) & 17.335 (0.050) & 16.541 (0.023) & 16.347 (0.032) & KAIT4 \\
2008 Feb 15 & 4512.18 & 175.43 & \nodata & 18.036 (0.030) & 17.466 (0.020) & 16.641 (0.023) & 16.229 (0.030) & Nickel \\
2008 Mar 4 & 4530.18 & 193.43 & \nodata & 18.353 (0.039) & 17.881 (0.025) & 16.971 (0.026) & 16.648 (0.030) & Nickel \\
2008 Nov 20 & 4791.15 & 454.40 & \nodata & \nodata & 21.577 (0.051) & \nodata & 21.250 (0.056) & HST \\
2008 Nov 25 & 4796.09 & 459.34 & \nodata & 22.221 (0.058) & \nodata & 21.388 (0.058) & \nodata & HST \\
\enddata
\tablenotetext{a}{$1\sigma$ uncertainties are given in parentheses.}
\tablenotetext{b}{Relative to $B$-band maximum (JD = 2,454,336.75).}
\tablenotetext{c}{KAIT4 stands for the fourth filter set used at KAIT; see Ganeshalingam et al. (2010) for details.}
\end{deluxetable}

\begin{deluxetable}{ccccc}
\tabletypesize{\scriptsize}
\tablecaption{Journal of Spectroscopic Observations of SN 2007gr.\label{tab:SPEC} }
\tablewidth{0pt}
\tablehead{
\colhead{Ut Date} & \colhead{JD-2,450,000} & \colhead{Phase\tablenotemark{a}} & \colhead{Range(\AA)}
& \colhead{Instrument\tablenotemark{b}}}
\startdata
2007 Aug 16.50 & 4329.00 & -7.8 & 3300--9580 & Kast\\
2007 Aug 17.63 & 4330.13 & -6.6 & 3000--9400 & LRISb\\
2007 Aug 21.51 & 4334.0 & -2.8 & 3300--10400 & Kast\\
2007 Sep 5.34 & 4348.84 & 12.1 & 3300--10400 & Kast\\
2007 Sep 19.43 & 4362.93 & 26.2 & 3300--10400 & Kast\\
2007 Oct 14.65 & 4388.15 & 51.4 & 3110--8712 & LRISb\\
2007 Nov 2.16 & 4406.66 & 69.9 & 3300--10400 & Kast\\
2007 Nov 10.44 & 4414.94 & 78.2 & 3300--10400 & Kast\\
2007 Dec 31.36 & 4465.86 & 129.1 & 3300--10400 & Kast\\
2008 Jan 15.32 & 4480.82 & 144.1 & 3300--10400 & Kast\\
2008 Feb 12.00 & 4509.00 & 172.3 & 3075--9340 & LRISb\\
2008 Aug 28.59 & 4707.09 & 370.3 & 3284--9248 & LRISb\\
2008 Oct 27.52 & 4767.02 & 430.2 & 3156--9170 & LRISb\\
\enddata
\tablenotetext{a}{Relative to $B$-band maximum (JD = 2,454,336.75).}
\tablenotetext{b}{Kast= Lick Shane 3\,m Kast spectrograph; LRISb = Keck-I 10\,m LRIS-blue.}
\end{deluxetable}

\begin{deluxetable}{cccc}
\tabletypesize{\scriptsize}
\tablecaption{Light-Curve Parameters of SN 2007gr.\label{tab:LCP} }
\tablewidth{0pt}
\tablehead{
\colhead{Band} & \colhead{$t_{\rm max}$} & \colhead{Peak obs mag.} & \colhead{Peak abs mag.}\\
\colhead{} & \colhead{$-2,454,000$} & \colhead{(mag)} & \colhead{(mag)}
}
\startdata
$U$ & $4334.7 \pm 1.3$ & $13.35 \pm 0.06$ & $-17.19 \pm 0.36$ \\
$B$ & $4336.8 \pm 0.6$ & $13.45 \pm 0.03$ & $-17.03 \pm 0.35$ \\
$V$ & $4339.3 \pm 0.5$ & $12.91 \pm 0.02$ & $-17.48 \pm 0.35$ \\
$R$ & $4340.4 \pm 0.7$ & $12.74 \pm 0.02$ & $-17.60 \pm 0.35$ \\
$I$ & $4341.6 \pm 0.6$ & $12.60 \pm 0.03$ & $-17.67 \pm 0.35$ \\
\enddata
\end{deluxetable}

\end{document}